\documentclass[aps,reprint]{revtex4-1}

\usepackage{amsmath}
\usepackage{amsfonts}
\usepackage{bbm}
\usepackage{graphicx}
\graphicspath{{./}{../graphics/}}
\usepackage{hyperref}

\newcommand{\dif}{{\rm d}}
\newcommand{\Z}{\mathbb{Z}}

\newcommand{\figIa}{fig1a.eps}
\newcommand{\figIb}{fig1b.eps}
\newcommand{\figII}{fig2.eps}

\begin{document}

\title{Consequences of moduli stabilization in the Einstein-Maxwell
  landscape} 

\author{C\'esar Asensio}
\email[]{casencha@unizar.es}
\author{Antonio Segu\'{\i}}
\email[]{segui@unizar.es}
\affiliation{Departamento de F\'{\i}sica Te\'orica, Universidad de Zaragoza}

\date{\today}

\begin{abstract}
  A toy landscape sector is introduced as a compactification of the
  Einstein-Maxwell model on a product of two-spheres.  Features of the
  model include:  moduli stabilization, a distribution of the
  effective cosmological constant of the dimensionally reduced 1+1
  spacetime, which is different from the analogous distribution of the
  Bousso-Polchinski landscape, and the absence of the so-called
  $\alpha^*$-problem.  This problem arises when the
  Kachru-Kallosh-Linde-Trivedi stabilization mechanism is naively
  applied to the states of the Bousso-Polchinski landscape.  The model
  also contains anthropic states, which can be readily constructed
  without needing any fine-tuning.
\end{abstract}


\maketitle

\section{Introduction}
\label{sec:intro}

The cosmological constant problem \cite{Weinberg:1988cp}, namely, the
smallness of the cosmological vacuum energy density when compared to
predictions of the Standard Model of particle physics, has been one of
the major problems faced by physicists over the last century.
Inflation \cite{Guth:2000ka} solved a plethora of classical problems
in cosmology, but the cosmological constant and coincidence problems
have remained. It is natural to look for a solution to these old
problems using the most powerful theory at our disposal, which at this
moment is string theory.  A striking feature of string theory is that
it can accommodate a huge number of vacuum solutions, collectively
known as the (string theory) Landscape
\cite{Bousso:2000xa,Susskind:2003kw}.  In a cosmological context, a
given state of the Landscape corresponds to a universe, and the
enormous number of universes in the Landscape is known as the
multiverse.  Of course, the vast majority of the universes in the
multiverse are very different from ours, and thus we need a
probability distribution on the multiverse in order to make
predictions.  The cosmological measure problem refers to the
difficulty in constructing such a probability distribution
unambiguously from first principles \cite{Vilenkin:2006xv}.

\subsection{The Bousso-Polchinski Landscape}
\label{sec:BP}

The entire Landscape is too complex to be readily modelled, but we
have comparatively simple models of it \cite{Denef:2008wq}.  Perhaps
the most explicit model is Bousso-Polchinski's \cite{Bousso:2000xa}
(BP), which has provided us with an elegant solution to the
cosmological constant problem.  In this setting, the states of the
Landscape are represented by the nodes of an integer lattice in
$J$-dimensional flux space, and the effective cosmological constant
$\lambda$ of a state labeled by integers $n_1,\cdots,n_J$ is given by
\begin{equation}
  \label{eq:1}
  \lambda = \Lambda + \frac{1}{2}\sum_{j=1}^J q_j^2n_j^2\,.
\end{equation}
In \eqref{eq:1}, $\Lambda$, a negative bare cosmological constant, and
the charges $\{q_j\}$ are parameters of the model.  For
incommensurable charges and large $J$, choices of integers $\{n_j\}$
are possible so that $\lambda$ can be made positive and very small.
This means that the BP Landscape contains states with an effective
cosmological constant as small as the observed value of our universe
$\lambda_{\mathrm{obs}}\approx 1.5\times10^{-123}$ (in units such that
$8\pi G = \hbar = c = 1$) \cite{Perlmutter:1998np,Riess:1998cb}
without the necessity of fine-tuning the parameters $\Lambda$,
$\{q_j\}$.

The most severe limitation of this model is the lack of a stability
analysis of the de Sitter states.  The first consequence is that the
parameters $\{q_j\}$ should be fixed \emph{a priori}.  Another
consequence is that the model identifies nodes in the lattice with
vacuum states of the theory.  The criteria for deciding if a lattice
node contains a state are the existence of a classical solution and
stability.  Unstable classical solutions cannot be counted among
physical states of the theory.  Therefore, a naive identification
between nodes and states will introduce many spurious vacua into the
model.

This has profound consequences on the predictions of the model.  The
measure problem previously mentioned makes it difficult to assign,
from first principles, a probability to each observable magnitude such
as the effective cosmological constant.  Thus, the first computations
are based only on abundances of states, which means that the \emph{a
  priori} probability distribution is uniform across the Landscape.
If we want to compute the probability $P_o = P(\lambda \approx
\lambda_{\mathrm{obs}})$, the answer requires the computation of the
quotient between the number of nodes satisfying the equality and the
total number of nodes in the Landscape.  This last number can be made
finite by means of a cutoff scale $\Lambda_{\mathrm{cutoff}}$ in flux
space, but then the desired probability is a negligibly small number.
Nevertheless, we also need a mechanism for populating the landscape,
such as eternal inflation \cite{Linde:1986fc}, resulting in a
dynamical reduction of the values of the cosmological constant
\cite{Bousso:2007er}.  Finally, it should be taken into account the
fact that we are interested only in those universes where
observer-hosting structures can develop, and the corresponding
anthropic probability distribution further modifies the prediction.
Therefore, the prior probability, the cosmological measure derived
from the population mechanism, and the anthropic factor are necessary
to accomplish a complete prediction of the emblematic probability
$P_o$.  Unfortunately, both dynamical relaxation and structure
formation probability distributions have a large support when compared
with $\lambda_{\mathrm{obs}}$, and thus the cosmological constant
problem is not completely solved by this model.  Other landscape
models with different prior probability distributions may dramatically
change the prediction, as recognized in \cite{SchwartzPerlov:2006hi}.

Thus, reliable predictions in the Landscape require a measure, but
also a complete characterization of the physical states of the system
by means of a stability analysis.  The Kachru-Kallosh-Linde-Trivedi
\cite{Kachru:2003aw} (KKLT) landscape model addressed this problem by
providing a mechanism for generating stable de Sitter (dS) states in a
landscape of supersymmetric and stable anti-de Sitter (AdS) vacua.
Unlike the BP landscape, there is only one stabilized modulus in this
model, and it is not straightforward to generalize the setting to a
large number of moduli.  Moreover, the lifting of AdS states to dS is
a quantum effect, and thus it is not completely clear if the stability
of AdS states is preserved in the process.  But even in the
affirmative case, no precise condition is given on the integers
labeling each different state beyond they being large.  AdS states are
stable for all physically acceptable integer configurations, but
stable dS states can have very restrictive conditions on the integers
labeling the nodes in the landscape.  Thus, preserving stability
unconditionally in the lifting is more than likely wrong.

\subsection{The $\alpha^*$-problem of the BP Landscape}
\label{sec:alpha-star}

One of the main advantages of the BP landscape is that the vacua
counting problems are often tractable, at least in an approximate
fashion.  For example, the distribution of the effective cosmological
constant values $\lambda$ can be approximately computed.  As some
authors anticipated \cite{SchwartzPerlov:2006hi}, the $\lambda$
distribution is flat near $\lambda=0$ \cite{Asensio:2008mq}.

There is another counting problem with a subtle consequence.  Let us
define the flux occupation number $\alpha$ as the fraction of nonzero
integers $n_1,\cdots,n_J$ of a given lattice node.  Assuming all
charges are equal, the probability distribution of the possible values
of $\alpha$ for the nodes inside a thin spherical shell around the
$\lambda=0$ value in flux space can be computed \cite{Asensio:2010fy}.
This distribution is approximately Gaussian with a peak located at a
value $\alpha^*$ which is less than one when $J$ is large.  The width
of the peak is of order $\frac{1}{\sqrt{J}}$, and thus we conclude
that the vast majority of the nodes inside the shell have a typical
value $J\alpha^*$ of nonzero integers.  This result is robust in the
sense that other sets in the Landscape yield the same probability
distribution.

As seen above, in the KKLT mechanism, the quantized fluxes of the
lifted states should be large in order to preserve its stability.  We
are forced to conclude that, if the two mechanisms are to be
reconciled, then the vast majority of the nodes of the BP landscape
will be unstable, and thus all counting problems, including the
emblematic probability $P_o$,
should be reconsidered.  This is what we have called the
\emph{$\alpha^*$-problem} of the BP landscape.

Now we may put the question, if a complete stability analysis in the
BP model were carried out, what would the effect of this new input on
$P_o$ be?  Perhaps the excluded states have very high $\lambda$ and
$P_o$ gets enhanced, or maybe the states contributing to $P_o$ have
some vanishing fluxes and $P_o$ becomes smaller, even zero.  It is
impossible to know in advance what will be the direction of the
modification.

\subsection{Motivation}
\label{sec:motivation}

We have seen above that the KKLT mechanism suggests that the vast
majority of the nodes in the BP landscape might have no associated
physical state, the $\alpha^*$-problem.  As far as we know, there is
currently no model combining the KKLT stabilization mechanism with the
BP solution of the cosmological constant problem.  Therefore, testing
the $\alpha^*$-problem requires finding a landscape toy model simple
enough to be exactly solvable, with many moduli to have a chance to
solving the cosmological constant problem, and having a detailed
characterization of the stable states.  The Einstein-Maxwell (EM)
landscape \cite{RandjbarDaemi:1982hi} can be compactified over a
product of two-spheres, the so-called multi-sphere Einstein-Maxwell
(MS-EM) landscape \cite{Asensio:2012pg}.  This model, described below,
fulfills these three requirements.  Thus, the motivation behind this
paper is to summarize the main properties of this landscape toy model,
interpreting the results as an indication of possible phenomena one
may encounter in more realistic models.  An exhaustive analysis of the
details of the model and its main consequences can be found in the
companion paper \cite{Asensio:2012pg}.

\section{The multi-sphere Einstein-Maxwell Landscape}
\label{sec:MS-EM}

The multi-sphere compactification of the EM model is
defined by the ansatz
\begin{equation}
  \label{eq:2}
  \dif s^2 = e^{2\phi(t,x)}\bigl(-\dif t^2 + \dif x^2\bigr)
  + \sum^J_{i=1} e^{2\psi_i(u_i,v_i)}\bigl(\dif u_i^2 + \dif v_i^2\bigr)\,.
\end{equation}
The metric (\ref{eq:2}) represents a manifold of the form
$\text{(A)dS}_2\times\bigl[\text{S}^2\bigr]^J$, which describes a
sector of the $2J+2$-dimensional EM theory, namely, the direct product
of a 1+1 cosmological solution and $J$ two-dimensional spheres.  Thus,
the moduli of the solution are the $J$ radii of the spheres.  The
exponents, $\phi(t,x)$ and $\psi_i(u_i,v_i)$, characterize conformal
representations of the $\mathrm{(A)dS}_2$ and $\mathrm{S}^2$ parts,
and thus they satisfy uncoupled Liouville equations
\begin{equation}
  \label{eq:3}
  \lambda = \bigl(\phi_{tt} - \phi_{xx}\bigr)e^{-2\phi}\,,\qquad
  K_i = -\Delta_i\psi_i\,e^{-2\psi_i}\,,
\end{equation}
where
\begin{itemize}
\item $\Delta_i$ is $i$-th Laplacian operator $\partial_{u_i}^2
  + \partial_{v_i}^2$.
\item $\lambda$ is the curvature of the $\mathrm{AdS}_2$ ($\lambda<0$)
  or $\mathrm{dS}_2$ part ($\lambda>0$), that is, the effective
  cosmological constant of the dimensionally reduced cosmology.
\item $K_i$ is the Gaussian curvature of the $i^{\mathrm{th}}$ sphere
  $S^2$.
\end{itemize}
In addition, the model also includes a bare, positive cosmological
constant $\Lambda$, which is the only parameter in the model, and an
electromagnetic field in a monopole-like configuration whose flux
through the $i^{\mathrm{th}}$ sphere is $Q_i$.  Dirac quantization
condition then reads $Q_{i}e = 2\pi n_{i}$, with $n_{i}\in\Z$, where
$e$ is the charge of test particles moving in the gravitational,
magnetic background.  We will absorb $e$ by redefining
$\frac{\Lambda}{e^2}\to\Lambda$, thereby rendering all magnitudes
dimensionless.

When inserted in the Einstein equation, the ansatz
(\ref{eq:2},\ref{eq:3}) produces an algebraic equation that $\lambda$
should satisfy, which depends on the node considered and on $\Lambda$.
This is the \emph{state existence} equation of a node:
\begin{equation}
  \label{eq:4}
  \Lambda = L_{n}(\lambda) \equiv 
  \frac{1}{2}\Bigl[J\lambda + \sum_{i=1}^J \frac{1}{n_i^2}
  \left(1 + s_i \sqrt{1 - 2\lambda\,n_i^2}\ \right)
  \Bigr]\,.
\end{equation}
In equation (\ref{eq:4}), the signs $s_i=\pm$ come from the solution
of a quadratic equation satisfied by the curvatures $K_{i}$ and give,
at least \emph{a priori}, several different equations for each given
node $n=(n_1,\cdots,n_J)$.  The equation obtained by setting all signs
to $+$ is called the \emph{principal branch}.  Each solution of
equation (\ref{eq:4}) for a given node $n$ is a possible state of the
MS-EM landscape.  Nevertheless, existence of a solution is not enough:
one must also demand positivity of all curvatures $K_i$ and reality of
$\lambda$.  This condition rules out negative signs in equation
(\ref{eq:4}) when looking for AdS states, because a single minus
produces a negative curvature.  Furthermore, positivity of square root
arguments in (\ref{eq:4}) implies the existence of a branching point
$\lambda_{\mathrm{b}} = \frac{1}{2\max_{1\le j\le J}\{n_j^2\}}$ in the
$L_{n}(\lambda)$ function, thus placing an upper limit on the values
of $\lambda$ which states can possibly have.

Therefore, states might exist if adequate solutions are found to the
existence equation (\ref{eq:4}), but they will be true physical
states only if they are stable.

Stabilization is addressed by perturbing the ansatz (\ref{eq:2}) to
\begin{equation}
  \label{eq:5}
  \dif s^2 =
  e^{2\phi - 2\sum_{i=1}^J \xi_i}
  \bigl(-\dif t^2 + \dif x^2\bigr) +
  \sum_{i=1}^J
  e^{2\psi_i + 2\xi_i} \bigl( \dif u_i^2 + \dif v_i^2\bigr)\,.  
\end{equation}
The perturbations $\xi_i(t,x)$ describe changes in the radii of the
internal spheres, and thus they will be called \emph{multi-radion}
fields.  In writing the equations of motion associated with the metric
ansatz (\ref{eq:5}) we insert (\ref{eq:3}) for the curvatures
$\lambda$, $K_i$ of the unperturbed solution, thus neglecting the
backreaction of the perturbations on the cosmological part.  After
linearizing the equations of motion about the unperturbed solution
$\xi_i=0$, \footnote{The effective action for the multi-radion field
  is a 1+1 theory with a Lagrangian where the radions are coupled by
  dilatonic factors, so that one cannot establish the stability of the
  $\xi_{i}=0$ solution by simply looking for minima of an explicit
  multi-radion potential.}  we obtain
\begin{equation}
  \label{eq:6}
  e^{-2\phi}\bigl[\partial_{tt}\boldsymbol{\xi}
  - \partial_{xx}\boldsymbol{\xi}\bigr] = -H\boldsymbol{\xi}\,,
\end{equation}
where $\boldsymbol{\xi}$ is the column vector of the radions and $H$
is a constant matrix formed out of a solution of (\ref{eq:4}).  The
stability criterion is therefore that the $H$ matrix should be
positive definite.

Some general stability results can be extracted from the
characteristic polynomial of $H$:
\begin{itemize}
\item All AdS states are stable.
\item All dS states having at least a vanishing flux number $n_i=0$
  are unstable.
\item All dS states coming from a non-principal branch are unstable.
  This leaves the principal branch of (\ref{eq:4}) as the only source
  of AdS and stable dS states.
\end{itemize}

Focusing on the principal branch of the existence equation, the
function $L_n(\lambda)$ has a maximum at $\lambda=0$, and thus two
solutions exist (one dS and another AdS) to the existence equation
$L_n(\lambda)=\Lambda$ near $\lambda=0$ if
\begin{equation}
  \label{eq:7}
  \sum_{i=1}^J \frac{1}{n_i^2} \ge \Lambda\,.
\end{equation}
We can see from (\ref{eq:4})
that Minkowski states with $\lambda=0$ can exist if and only if
equality is satisfied in (\ref{eq:7}).  The corresponding equation
\begin{equation}
  \label{eq:8}
  \sum_{i=1}^J \frac{1}{n_i^2} = \Lambda\,.
\end{equation}
defines a null-$\lambda$ hypersurface in flux space separating dS from
AdS states on the principal branch.  This hypersurface has asymptotic
hyperplanes given by $|n_i| = \frac{1}{\sqrt{\Lambda}}$, and no dS
state can exist below this value because of the branching point (AdS
states should only obey inequality (\ref{eq:7})).  Thus, all dS
states are confined between the null-$\lambda$ hypersurface
(\ref{eq:8}) and its asymptotic hyperplanes, so that flux numbers in
a node cannot be smaller than $\frac{1}{\sqrt{\Lambda}}$; otherwise a
dS state will not exist at such a node.

Immediately one concludes that all nodes near the coordinate
hyperplanes are devoid of states, and thus the $\alpha^{*}$-problem is
absent in the MS-EM landscape.

\begin{figure}
  \centering
  \includegraphics{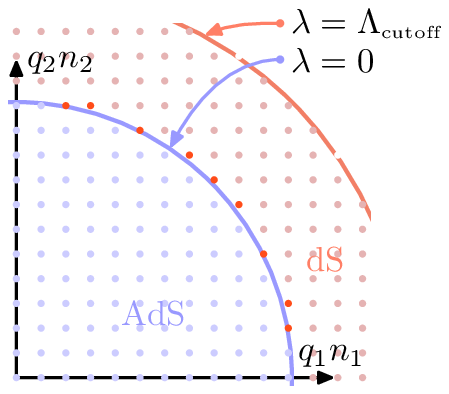}\hfill
  \includegraphics{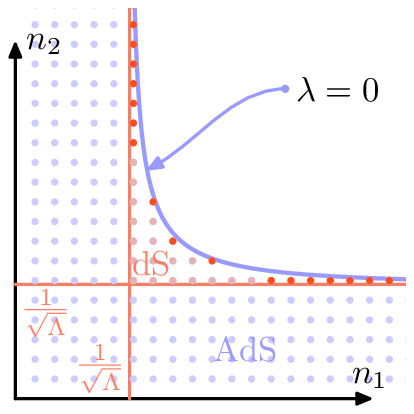}%
  \caption{\label{fig:landscapes} Left panel:  Example of a $J=2$ BP
    landscape, showing the null-$\lambda$ curve separating dS from AdS
    states, some random low-$\lambda$ states (red) and the cutoff
    curve.  Right panel: a $J=2$ MS-EM landscape, showing
    the null-$\lambda$ curve separating dS and AdS states (shown
    superimposed as they are different solutions of (\ref{eq:4})),
    its asymptotes, some random low-$\lambda$ states and state chains
    (red).}%
\end{figure}

In the BP landscape, the null-$\lambda$ hypersurface is a sphere, as
can be seen from equation (\ref{eq:1}).  In contrast, the
null-$\lambda$ hypersurface (\ref{eq:8}) in the MS-EM
landscape is not compact, and this allows the existence of \emph{state
  chains}, see figure \ref{fig:landscapes} (right) for a example in
the $J=2$ case.

\begin{figure}
  \centering
  \includegraphics{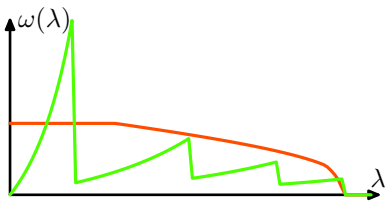}
  \caption{\label{fig:distributions}Effective cosmological constant
    distribution of a $J=2$ BP landscape (flat curve), and a $J=2$
    MS-EM landscape (jagged curve).}%
\end{figure}

Chained states are arranged by decreasing $\lambda$, and the states
with lowest $\lambda$ are always stable.  They contribute to the
effective cosmological constant distribution in peaks, which become
very sharp when $\frac{1}{\sqrt{\Lambda}}$ approaches an integer from
below.  Thus, the $\lambda$ distribution has a dominant peak coming
from the longest state chains, and subdominant peaks separated by a
gap from the dominant one which merge with a bulk distribution having
an almost constant average behavior before vanishing after reaching
the maximum $\lambda$ value of stable dS states, which is
$\lambda_{\mathrm{max}} = \frac{2\Lambda}{J(J+3)}$.  Figure
\ref{fig:distributions} summarizes the very different behavior of both
distributions.

Therefore, the special form of the null-$\lambda$ hypersurface
(\ref{eq:8}) leads to state chains, which generate peaks in the
low-$\lambda$ region of the $\lambda$ distribution, thereby providing
an alternative mechanism for finding small values of the effective
cosmological constant besides the random closeness, which is also
present.

\section{Anthropic states in the MS-EM landscape}
\label{sec:anthropic}

We can look for a set of integers $\{n_1,\cdots,n_J\}$ which
approximately solve equation (\ref{eq:8}), expecting that reasonably
good choices will yield very low values of the 1+1 cosmological
constant $\lambda$.  Choosing the best integer step by step we arrive
at the following recurrence relation:
\begin{equation}
  \label{eq:9}
  \Lambda_{j+1} = f(\Lambda_j) = \Lambda_j - 
  \left\lceil \Lambda_j^{-\frac{1}{2}}\right\rceil^{-2}
  \,,\quad
  n_j = \left\lceil\Lambda_j^{-\frac{1}{2}}\right\rceil\,.
\end{equation}
The initial value triggering the recurrence is $\Lambda_1=\Lambda$.
After $J$ steps, we obtain a solution choosing the last integer as
$n_J = \bigl\lfloor\Lambda_J^{-1/2}\bigr\rfloor$.  Equation
(\ref{eq:9}) is a fixed point iteration with superlinear convergence
rate, whose solution is thus a double exponential
$ \Lambda_j \approx 2^{\sum^{j-2}_{k=0}(\frac{3}{2})^k}
\Lambda^{(\frac{3}{2})^{j-1}},\ {}
(j\ge2)$.  It can be shown \cite{Asensio:2012pg} that the resulting
fast-growing integers $\{n_j\}$ form a node which has always a
well-defined stable state on it.  Moreover, this node is the end of a
very long state chain, which translates in a very narrow peak in the
$\lambda$ distribution containing $\sim n_J$ states whose support is
the interval $[0,\Lambda_J/2]$.  Thus, we can find the whole peak
inside the anthropic range $0\le\lambda\le\lambda_{\mathrm{A}}$ for a
given $\Lambda$ by equating $\lambda_{\mathrm{A}} = \Lambda_J/2$ and
solving for $J$, resulting in $ J = 1 +
\log_{\frac{3}{2}}\Bigl(\frac{\log(\lambda_\mathrm{A}/2)}{\log(4\Lambda)}\Bigr)$.
As an example, with $\lambda_{\mathrm{A}}=10^{-120}$, we can obtain an
anthropic peak using $\Lambda=0.1$ and $J=15$, yielding $\sim10^{57}$
states.  Using $\Lambda=0.0002$ and $J=10$ we obtain $10^{59}$
anthropic states with the same $\lambda_\mathrm{A}$.  We can see that
the MS-EM landscape contains a huge amount of anthropic states with
moderate values of $J$ for any $\Lambda$, and thus no fine-tuning is
needed.

It can be seen that states in the anthropic chains just described are
non-generic despite being very numerous.  Nevertheless, the peak in
the prior distribution can be made very narrow when compared with the
full anthropic range, and thus the anthropic factor influencing the
cosmological constant prediction can be considered as almost constant.
As emphasized in \cite{SchwartzPerlov:2006hi}, the form of the prior
distribution can completely change the prediction, and the narrow peak
provided by anthropic chains is an example where the prior can
dominate the prediction of the cosmological constant's value.

\section{Conclusions}
\label{sec:conclusions}

Trying to reconcile the BP landscape with the KKLT stabilization
mechanism leads to the $\alpha^*$-problem of the BP model.  Addressing
this problem requires a model where an exact solution of stable dS and
AdS states can be found, and a very simple example of this model is
given by the MS-EM landscape.  Looking at the states found in this
model, we can extrapolate that the assumed fixing of the moduli in the
BP case might not be enough to guarantee the existence and stability
of the states in all nodes.  Thus, we should conclude that such an
analysis would dramatically change the conclusions of all counting
problems in the BP landscape, in particular the predictions concerning
the number of anthropic states.

Moreover, the non-trivial geometrical features of the null-$\lambda$
hypersurface of the MS-EM landscape lead to the existence of state
chains, which provide a new mechanism for finding low-$\lambda$
states.  This would largely affect all probability computations in
this context, as reflected by the existence of a huge number of
anthropic states in the model.  This non-trivial geometrical fact,
with such profound implications in the predictions of the theory, may
also well be present, maybe under different forms, in the true string
theory landscape.

\begin{acknowledgments}
  We would like to thank Noel Hughes and Susana Gonz\'alez for reading
  this manuscript, and the Pedro Pascual Benasque Center of Science.
  We also thank F.~Denef, R.~Emparan, J.~Garriga, B.~Janssen,
  D.~Marolf and J.~Zanelli for useful discussions and encouragement.
  This work has been supported by CICYT (grant FPA-2009-09638) and
  DGIID-DGA (grant 2011-E24/2).  We also thank the support given by
  grant A9335/10.
\end{acknowledgments}


%

\end{document}